\documentclass[10pt,conference]{IEEEtran}

\usepackage{cite, amsmath, amssymb, enumerate, units, empheq, graphicx, subfig, comment, url}

\newtheorem{assum}{Assumption}
\newtheorem{lemma}{Lemma}

\newtheorem{theorem}{Theorem}
\newtheorem{prop}{Proposition}

\newcommand{\ourdom}{\mathbb{R}^d}
\newcommand{\extRp}{(-\infty, \infty]}
\newcommand{\dom}{\mbox{dom}}

\newcommand{\oo}[1]{\frac{1}{#1}}
\newcommand{\argmin}{\operatornamewithlimits{argmin}}
\newcommand{\norm}[1]{{\left\| #1 \right\|}}
\newcommand{\ak}{\alpha}

\newcommand{\inn}[2]{\left \langle #1, #2 \right \rangle}
\newcommand{\sumkn}{\sum \limits_{k=1}^n}

\newcommand{\sumi}{\sum_{i=1}^m}
\newcommand{\sumj}{\sum_{j=1}^m}
\newcommand{\sumk}{\sum_{k=1}^\infty}
\newcommand{\sumkz}{\sum_{k=0}^\infty}

\newcommand{\paren}[1]{\left( #1 \right) }
\newcommand{\brac}[1]{\left[ #1 \right] }

\newcommand{\xik}{x_i^{(k)}}
\newcommand{\xjk}{x_j^{(k)}}
\newcommand{\xikm}{x_i^{(k-1)}}
\newcommand{\xjkm}{x_j^{(k-1)}}
\newcommand{\xbar}[1]{\overline{x}^{(#1)}}

\newcommand{\yik}{y_i^{( k )}}
\newcommand{\yjk}{y_j^{( k )}}
\newcommand{\yikm}{y_i^{( k-1 )}}
\newcommand{\yjkm}{y_j^{( k-1 )}}
\newcommand{\ybar}[1]{\overline{y}^{(#1)}}

\newcommand{\wik}{w_i^{( k )}}
\newcommand{\wikm}{w_i^{( k-1 )}}

\newcommand{\zik}{z_i^{( k )}}
\newcommand{\qik}{q_i^{( k )}}
\newcommand{\qjk}{q_j^{( k )}}
\newcommand{\qikm}{q_i^{( k-1 )}}
\newcommand{\qjkm}{q_j^{( k-1 )}}
\newcommand{\qhik}{\hat{q}_i^{( k )}}
\newcommand{\qhikm}{\hat{q}_i^{( k-1 )}}
\newcommand{\qhjkm}{\hat{q}_j^{( k-1 )}}
\newcommand{\qbar}[1]{\overline{q}^{(#1)}}

\newcommand{\lijk}{\lambda_{ij}^{( k )}}

\newcommand{\lijkhatm}{\hat{\lambda}_{ij}^{( k-1 )}}
\newcommand{\betak}{\frac{k-1}{k+2}}
\newcommand{\betakm}{\beta_{k-1}}
\newcommand{\ek}{e^{( k )}}
\newcommand{\epsk}{\varepsilon^{( k )}}

\newcommand{\prox}{\mbox{prox}}
\newcommand{\proxh}[1]{\prox_h^\ak \left\{ #1 \right\} }

\IEEEoverridecommandlockouts

\begin{document}

\title{\vspace{0.5in} A Fast Distributed Proximal-Gradient Method}

\author{
Annie I. Chen and Asuman Ozdaglar
\thanks{This work was supported by National Science Foundation under Career grant DMI-0545910 and AFOSR MURI FA9550-09-1-0538.}
\thanks{The authors are with the Laboratory for Information and Decision Systems, Massachusetts Institute of Technology, 77 Massachusetts Ave, Cambridge, MA 02139. \texttt{ \{anniecia, asuman\}@mit.edu}}
}

\markright{}

\maketitle

\thispagestyle{headings}

\begin{abstract}
We present a distributed proximal-gradient method for optimizing the average of convex functions, each of which is the private local objective of an agent in a network with time-varying topology. The local objectives have distinct differentiable components, but they share a common nondifferentiable component, which has a favorable structure suitable for effective computation of the proximal operator. In our method, each agent iteratively updates its estimate of the global minimum by optimizing its local objective function, and exchanging estimates with others via communication in the network. Using Nesterov-type acceleration techniques and multiple communication steps per iteration, we show that this method converges at the rate $1/k$ (where $k$ is the number of communication rounds between the agents), which is faster than the convergence rate of the existing distributed methods for solving this problem. The superior convergence rate of our method is also verified by numerical experiments. 
\end{abstract}

\section{Introduction}

There has been a growing interest in developing distributed methods that enable the collection, storage, and processing of data using multiple agents connected through a network. Many of these problems can be formulated as 
\begin{equation}
\min_{x \in \ourdom} f(x) = \oo{m} \sumi f_i(x), \label{optproblem}
\end{equation}
where $m$ is the number of agents in the network, $f(x)$ is the global objective function, and for each $i=1,\ldots,m$, $f_i(x)$ is a local objective function determined by private information available to agent $i$. The goal is for agents to cooperatively solve problem (\ref{optproblem}). Most methods for solving this problem involve each agent maintaining an estimate of the global optimum of problem (\ref{optproblem}) and updating this estimate iteratively using his own private information and information exchanged with neighbors over the network. Examples include a team of sensors exploring an unknown terrain, where $f_i(x)$ may represent regularized least-squares fit to the measurement taken at agent $i$. As another example, in a distributed machine learning problem, $f_i(x)$ may represent a regularized loss function according to training samples accessible to agent $i$.

Most optimization algorithms developed for solving problem (\ref{optproblem}) and its variations are first-order methods (i.e., methods that use gradient or subgradient information of the objective functions), which are computationally inexpensive and naturally lead to distributed implementations over networks. These methods typically {\it converge at rate $1/\sqrt{n}$}, where $n$ is the number of communication steps in which agents exchange their estimates; in other words,  the difference between the global objective function value at an agent estimate and the optimal value of problem (\ref{optproblem}) is inversely proportional to the square-root of the number of communication steps carried out (see \cite{NeO09} for a distributed subgradient method and \cite{DAW11} for a distributed dual averaging algorithm with this rate). An exception is the recent independent work \cite{JXM11}, which developed a distributed gradient method with a diminishing step size rule, and showed that under certain conditions on the communication network and higher-order differentiability assumptions, the method converges at rate $\log(n)/n$. 

In this paper, we focus on a structured version of problem (\ref{optproblem}) where the local objective function $f_i(x)$ takes the additive form $f_i(x)=g_i(x)+h(x)$ with $g_i$ a differentiable function and $h$ a common nondifferentiable function.\footnote{This models problems in which $h$ represents a nondifferentiable regularization term. For example, a common choice for machine learning applications is $h(x) = \lambda \norm{x}_1$, where $\norm{x}_1$ is the sum of the absolute values of each element of the vector $x$.}  We develop a distributed proximal gradient method that solves this problem {\it at rate $1/n$ over a network with time-varying connectivity}. Our method involves each agent maintaining an estimate of the optimal solution to problem (\ref{optproblem}) and updating it through the following steps: at iteration $k$, each agent $i$ takes a step along the negative gradient direction of $g_i$, the differentable component of his local objective function, and then enters the {\it consensus stage} to exchange his estimate with his neighbors. The consensus stage consists of $k$ {\it communication steps}. In each communication step, the agent updates his estimate to a linear combination of his current estimate and the estimates received from neighbors. After the consensus stage, each agent performs a proximal step with respect to $h$, the nondifferentiable part of his objective function, at his current estimate, followed by a Nesterov-type acceleration step.

This algorithm has two novel features: first, the multi-step consensus stage brings the estimates of the agents close together before performing the proximal step, hence allowing us to reformulate this method as an inexact centralized proximal gradient method with controlled error. Our analysis then uses the recent results on the convergence rate of an inexact (centralized) proximal-gradient method (see \cite{SRB11}) to establish the convergence rate of the distributed method.  Second, exploiting the special structure in the objective functions allows for the use of a proximal-gradient method that can be accelerated using a Nesterov acceleration step, leading to the faster convergence rate of the algorithm.

Other than the papers cited above, our paper is related to the seminal works \cite{Tsi84} and \cite{TBA86}, which developed distributed methods for solving global optimization problems with a common objective (i.e., $f_i(x)=f(x)$ for all $i$ in problem (\ref{optproblem})) using parallel computations in multiple servers. It is also related to  the literature on consensus problems and algorithms (see \cite{Jad03, Olf04, Boy05, Ols06}) and a recent growing literature on multi-agent optimization where information is decentralized among multiple agents connected through a network (see \cite{Joh08, Ram09, NOP10, LoO11, Zhu12} for subgradient algorithms and \cite{Boy10, Wei12} for algorithms based on the alternating direction method of multipliers). Finally, our paper builds on the seminal papers on (centralized) proximal-point and proximal-gradient methods (see \cite{Roc76, BeT09, BeT10}).


The paper is organized as follows: 
Section 2 describes preliminary results pertinent to our work. 
In Section 3, we introduce our fast distributed proximal-gradient method and establish its convergence rate. 
Section 4 presents numerical experiments that verify the effectiveness of our method. 
Finally, Section 5 concludes the paper with open questions for future work. 

\noindent {\bf Notations and Definitions:} \vskip .5pc

\begin{itemize}
\item 
For a vector or scalar that is local, we use subscript(s) to denote the agent(s) it belongs to, and superscripts with parentheses to denote the iteration number; for example, $\xik$ denotes the estimate of agent $i$ at iteration $k$. 

\item 
For a vector or scalar that is common to every agent, or is part of a centralized formulation, the iteration number is also written in superscripts with parentheses; for example, $\xbar{k}$ denotes the average estimate of all agents at iteration $k$. Similarly, $\ek$ and $\epsk$ denote the errors in the centralized formulation at iteration $k$.

\item 
The standard inner product of two vectors $x,y \in \mathbb{R}^d$ is denoted $\inn{x}{y} = x'y$. For $x \in \mathbb{R}^d$, its Euclidean norm is $\norm{x} = \sqrt{\inn{x}{x}}$, and its 1-norm is $\norm{x}_1 = \sum_{l=1}^d |x(l)|$, where $x(l)$ is its $l$-th entry.

\item 
For a matrix $A$, we denote its entry at the $i$-th row and $j$-th column as $[A]_{ij}$. We also write $[a_{ij}]$ to represent a matrix $A$ with $[A]_{ij} = a_{ij}$. A matrix $A$ is said to be \emph{stochastic} if the entries in each row sum up to $1$, and it is \emph{doubly stochastic} if $A$ and $A'$ are both stochastic.

\item We write $a(n) = O(b(n))$ if and only if there exists a positive real number $M$ and a real number $n_0$ such that $|a(n)| \leq M|b(n)|$ for all $n \geq n_0$.

\item For a function $F: \ourdom \rightarrow (-\infty,\infty]$, we denote the domain of
$F$ by $\dom(F)$, where
\[\dom(F)=\{x\in\ourdom \mid F(x)<\infty\}.\]
For a given vector $x \in \dom(F)$, we say that $z_F(x)\in\ourdom$ is a \emph{subgradient} of the function $F$ at $x$ when the following relation holds:
\[ F(x) + \inn{z_F(x)}{y-x} \leq F(y)\qquad \hbox{for all } x\in\dom(F). \]
The set of all subgradients of $F$ at $x$ is denoted by $\partial F(x)$.

\end{itemize}

\section{Preliminaries}

In this section, we introduce the main concepts and establish key results on which our subsequent analysis relies. 
Section 2.1 gives properties of the proximal operator; Section 2.2 summarizes convergence rate results for an inexact centralized proximal-gradient method characterized in terms of the errors introduced in the method.

\subsection{Properties of the Proximal Operator}

For a closed proper convex function $h: \mathbb{R}^d \rightarrow (-\infty, \infty]$ and a scalar $\alpha>0$, we define the {\it proximal operator with respect to $h$} as
\[\prox_h^\alpha \{x\} = \argmin_{z \in \ourdom}\left\{ h(z) + \oo{2 \alpha} \norm{z-x}^2  \right\}.\]
It follows that the minimization in the preceding optimization problem is attained at a unique point $y = \prox_h^\alpha\{x\}$, i.e., the proximal operator is a single-valued map \cite{Roc76}. Moreover, using the optimality condition for this problem
\[0 \in \partial h(y) + \oo{\alpha}(y-x),\] 
we can see that the proximal operator has the following properties \cite{BeT09}:

\begin{prop} (Basic properties of the proximal operator) \label{p_prox}
Let $h: \mathbb{R}^d \rightarrow (-\infty, \infty]$ be a closed proper convex function. For a scalar $\alpha>0$ and $x\in \mathbb{R}^d$, let $y = \prox_h^\alpha \{x\}$.
\begin{enumerate} [(a)] 
\item We have $\oo{\alpha}(x-y) \in \partial h(y)$.
\item The vector $y$ can be written as $y = x - \alpha z$, where $z \in \partial h(y)$.
\item We have $h(u) \geq h(y) + \oo{\alpha}\inn{x-y}{u-y}$ for all $u \in \ourdom$.
\item (Nonexpansiveness) For $x, \hat{x}\in \mathbb{R}^d$, we have
\[\norm{\prox_h^\alpha \{x\} - \prox_h^\alpha\{ \hat{x} \}} \leq \norm{x-\hat{x}}.\]
\end{enumerate}
\end{prop}

\subsection{Inexact Proximal-Gradient Method}

Our approach for the analysis of the proposed distributed method is to view it as an inexact centralized proximal-gradient method, with the error controlled by multiple communication steps at each iteration. This allows us to use recent results on the convergence rate of an inexact centralized proximal-gradient method to establish the convergence rate of our distributed method.
The following proposition from \cite{SRB11} characterizes the convergence rate of an inexact proximal-point method in terms of error sequences $\{\ek\}_{k=1}^\infty$ and $\{\epsk\}_{k=1}^\infty$.

\begin{prop} \label{l_inexact}
\cite[Proposition 2]{SRB11}
Let $g: \ourdom \rightarrow \mathbb{R}$ be a convex function that has a Lipschitz continuous gradient with Lipschitz constant $L$, and let $h: \ourdom \rightarrow \extRp$ be a lower semi-continuous proper convex function. Suppose the function $f = g+h$ attains its minimum at a certain $x^* \in \ourdom$. 

Given two sequences $\{\ek\}_{k=1}^\infty$ and $\{\epsk\}_{k=1}^\infty$, where $\ek \in \ourdom$ and $\varepsilon \in \mathbb{R}$ for every $k$, consider the accelerated inexact proximal gradient method, which iterates the following recursion:
\begin{align} \label{inexact}
\begin{cases}
x^{( k )} &\in \prox_{h, \epsk}^\alpha \{ y^{( k-1 )} - \alpha \paren{ \nabla g(y^{( k-1 )}) + \ek } \} \\
y^{( k )} &= x^{( k )} + \frac{k-1}{k+2} \paren{ x^{( k )} - x^{( k-1 )} }
\end{cases}
\end{align}
where the step size is $\alpha = \oo{L}$, and 
\begin{align} \label{eprox}
&\prox_{h, \varepsilon}^\alpha\{y\} = \bigl\{ x \in \ourdom \mid \nonumber \\
&h(x) + \oo{2 \alpha}\norm{x-y}^2 \leq \min_{z \in \ourdom}\paren{h(z) + \oo{2 \alpha}\norm{z-y}^2 } + \varepsilon \bigr\}
\end{align}
indicates the set of all $\varepsilon$-optimal solutions for the proximal operator.

Then, for all $n \geq 1$, we have
\[f(x(n)) - f(x^*) \leq \frac{2L \paren{\norm{x^{( 0 )} - x^*} + 2 \tilde{A}_n + \sqrt{2 \tilde{B}_n} }^2}{(n+1)^2} \]
where
\[\tilde{A}_n = \sumkn k \paren{ \frac{\norm{\ek}}{L} + \sqrt{\frac{2 \epsk}{L}} }, \quad
\tilde{B}_n = \sumkn \frac{k^2 \epsk}{L}.\]
\end{prop}

Proposition \ref{l_inexact} indicates that as long as the error sequences $\{\norm{\ek}\}_{k=1}^\infty$ and $\{\epsk\}_{k=1}^\infty$  are such that the sequences $\{k \norm{\ek}\}_{k=1}^\infty$ and $\{k \sqrt{\epsk}\}_{k=1}^\infty$ are both summable, then the accelerated inexact gradient method achieves the optimal convergence rate of $O(\oo{n^2})$. 
It is straightforward to verify using the analysis in \cite{SRB11} that the result also holds for a constant step size $\alpha \leq \oo{L}$.


We shall see that error sequences in our inexact formulation, introduced by the distributed nature of our problem and controlled by multi-step consensus, can be bounded by sequences of the form $\{p^{( k )} \gamma^k\}_{k=1}^\infty$ for some polynomial $p^{( k )}$ of $k$ and some $\gamma \in (0,1)$, which we shall henceforth refer to as \emph{polynomial-geometric sequences}.  The next proposition shows that such sequences are summable and allows us to use Proposition \ref{l_inexact} in the convergence analysis of our method (the proof is omitted due to limited space). 

\begin{prop} \label{l_polygeo}
(Summability of polynomial-geometric sequences) 

Let $\gamma$ be a positive scalar such that $\gamma < 1$, and let \[P(k,N) = \{c_N k^N + ... + c_1 k + c_0 \mid c_j \in \mathbb{R}, j = 0,...,N \}\] denote the set of all $N$-th order polynomials of $k$, where $N$ is a nonnegative integer. Then for every polynomial $p^{( k )} \in P(k,N),$
\[\sumkz p^{( k )} \gamma^k < \infty.\]

\end{prop}

The result of this proposition for $p(k,N) = k^N$ will be particularly useful for our analysis in the upcoming sections. Therefore, we make the following definition:
\begin{align}	\label{sngamma}
S_N^\gamma := \sumkz k^N \gamma^k < \infty.
\end{align}

\section{Model and Method}

We consider the optimization problem
\begin{equation} \min_{x \in \ourdom} f(x) := \oo{m} \sumi f_i(x),\label{globalprob} \end{equation}
where $f(x)$ is the global objective function, and $f_i(x) = g_i(x) + h(x), i=1,..,m$ are local objective functions that are private to each agent. For example, for regularized logistic regression, the local objective functions are given by $g_i(x) = \oo{|N_i|} \sum_{j \in N_i} \log \paren{1+\exp (-b_j \inn{a_j}{x})}$ and $h(x) = \lambda \norm{x}_1$, where $N_i$ is the training dataset of agent $i$, corresponding to $\{a_j \mid j \in N_i\}$, the set of feature vectors, and $\{b_j \mid j \in N_i\}$, the set of associated labels.

We adopt the following assumption on the functions $g_i(x)$ and $h(x)$.

\begin{assum} \label{a_func} 

\begin{enumerate}[(a)]
\item For every $i$, $g_i: \ourdom \rightarrow \mathbb{R}$ is convex, continuously differentiable, and has a Lipschitz-continuous gradient with Lipschitz constant $L>0$, i.e., \[\norm{\nabla g_i(x) - \nabla g_i(y)} \leq L \norm{x-y}\qquad \hbox{for all }x,y \in \ourdom.\]
\item There exists a scalar $G_g$ such that for every $i$ and for every $x \in \ourdom$, $\norm{\nabla g_i(x)} < G_g$.
\item $h: \ourdom \rightarrow \mathbb{R}$ is convex. 
\item There exists a scalar $G_h$ such that for every $x \in \ourdom$, $\norm{z} < G_h$ for each subgradient $z \in \partial h(x)$.\item $f(x) = \oo{m} \sumi f_i(x) = \oo{m} \sumi g_i(x) + h_i(x)$ attains its minimum at a certain $x^*$. 
\end{enumerate}
\end{assum}

These assumptions are standard in the analysis of distributed first-order methods (see \cite{TBA86}, \cite{BHO05} and \cite{NeO09}).

We propose the following {\it distributed proximal-gradient method} for solving problem (\ref{globalprob}): Starting from initial estimates $\{y_i^{( 0 )}\}_{i=1,\ldots,m}$ with $y_i^{( 0 )}\in \mathbb{R}^d$, each agent $i$ updates his estimate $y_i^{( k-1 )}$ at iteration $k$ as follows:
\begin{subequations} \label{ch4_pg}
\begin{empheq}[left=\empheqlbrace]{align} 
\qik &= \yikm - \alpha \nabla g_i(\yikm) \label{ch4_pg_1}
\\
\qhik &= \sumj \lijk \qjk \label{ch4_pg_2}
\\
\xik &= \proxh{ \qhik } \label{ch4_pg_3}
\\
\yik &= \xik + \betak \paren{\xik - \xikm} \label{ch4_pg_4}
\end{empheq}
\end{subequations}

Here, $\alpha>0$ is a constant stepsize which is also the constant scalar used in the proximal operator. The scalars $\lambda_{ij}^{( k )}$ are weights given by
\[\lijk = [\Phi(t^{( k )}+k, t^{( k )})]_{ij},\]
for all $i,j=1\ldots,m$ and all $k \ge s$,
where $t^{( k )}$ is the total number of communication steps before iteration $k$, and $\Phi$ is a {\it transition matrix} representing the product of matrices $A(t)$, i.e., for $t\geq 0$,
\[\Phi(t,s) = A(t)A(t-1) \cdots A(s+1)A(s),\]
where $A(t)=[a_{ij}(t)]_{i,j=1,\ldots,m}$ is a matrix of weights $a_{ij}(t)\ge 0$ for $i,j=1,\ldots,m$. Using a vector notation $\hat{q}^{( k )}=[\hat{q}_i^{( k )}]_{i=1,\ldots,m}$ and $q^{( k )}=[q_i^{( k )}]_{i=1,\ldots,m}$, we can rewrite \eqref{ch4_pg_2} as
\begin{align*}
\hat{q}^{( k )} 
&= \Phi(t^{( k )}+k, t^{( k )}) q^{( k )} \\
&= A(t^{( k )}+k)A(t^{( k )}+k-1)\cdots A(t^{( k )}) q^{( k )}.
\end{align*}
Hence, this step represents agents performing $k$ communication steps at iteration $k$. At each communication step, agents exchange their values $q_i^{( k )}$ and update these values by linearly combining the received values using weights $A(t)$. We refer to \eqref{ch4_pg_2} as a {\it multi-step consensus stage}, since linear (in fact, convex, as we shall see) combinations of estimates will serve to bring the agent estimates close to each other. 

Our method involves each agent updating his estimate along the negative gradient of the differentiable part of his local objective function (step \eqref{ch4_pg_1}), a multi-step consensus stage (step \eqref{ch4_pg_2}), and a proximal step with respect to the nondifferentiable part of his local objective function (step  \eqref{ch4_pg_3}), which is then followed by a Nesterov-type acceleration step (step \eqref{ch4_pg_4}). Hence, it is a distributed proximal-gradient method with a multi-step consensus stage inserted before the proximal step. The multi-step consensus stage serves to bring the estimates close to each other before performing the proximal step with respect to the nondifferentiable function $h$. This enables us to control the error in the reformulation of the method as an inexact 
centralized proximal-gradient method.

We analyze the convergence behavior of this method under the information exchange model developed in \cite{Tsi84,NeO09}, which we summarize in this section. Let $A(t) = [a_{ij}(t)]$ be the weight matrix used in communication step $t$ of the consensus stage.
While the weight matrix $A(t)$ may be time-varying, we assume that it satisfies the following conditions for all $t$.
\begin{assum} \label{a_network} 
(Weight Matrix and Network Conditions)

Consider the weight matrices $A(t) = [a_{ij}(t)], t = 1,2, \ldots$.
\begin{enumerate}[(a)] 
\item (Double stochasticity) For every $t$, $A(t)$ is doubly stochastic.
\item (Significant weights)  There exists a scalar $\eta \in (0,1)$ such that for all $i$, $a_{ii}(t) \geq \eta$, and for $j \neq i$, either $a_{ij}(t) \geq \eta$, in which case $j$ is said to be a neighbor of $i$, and receives the estimate of $i$, at time $t$; or $a_{ij}(t) = 0$, in which case $j$ is not a neighbor of $i$ at time $t$.
\item (Connectivity and bounded intercommunication intervals)
Let 
\begin{align*}
E_t =& \{(j,i) \mid j \mbox{ receives the estimate of $i$ at time }t\}, \\
E_\infty =& \{(j,i) \mid j \mbox{ receives the estimate of $i$ for infinitely } \\ & \mbox{ many } t\}.
\end{align*}
Then $E_\infty$ is connected. 
Moreover, there exists an integer $B \geq 1$ such that if $(j,i) \in E_\infty$, then 
$(j,i) \in E_t \cup E_{t+1} \cup ... \cup E_{t+B-1}$.
\end{enumerate}
\end{assum}

In this assumption, part (a) ensures that each agent's estimate exerts an equal influence on the estimates of others in the network. Part (b) guarantees that in updating his estimate, each agent gives significant weight to his current estimate and the estimates received from his neighbors.  Part (c) states that the overall communication network is capable of exchanging information between any pair of agents in bounded time. An important implication of this assumption is that for $\overline{B} = (m-1)B$, then $E_t \cup E_{t+1} \cup ... \cup E_{t+\overline{B}-1} = E_\infty$.

The following result from \cite{NeO09} on the limiting behavior of products of weight matrices will be key in establishing the convergence of our algorithm in the subsequent analysis.
 
\begin{prop} \label{l_cons}
\cite[Proposition 1(b)]{NeO09}
Let Assumption 2 hold, and for $t \geq s$, let
\[\Phi(t,s) = A(t)A(t-1) \cdots A(s+1)A(s).\]
Then the entries $[\Phi(t,s)]_{ij}$ converges to $\oo{m}$ as $t \rightarrow \infty$ with a geometric rate uniformly with respect to $i,j$. Specifically, for all $i,j \in \{1,...,m\}$ and all $t,s$ with $t \geq s$,
\[ \left| [\Phi(t,s)]_{ij} - \oo{m} \right| \leq 2 \frac{1+\eta^{-\overline{B}}}{1-\eta^{\overline{B}}} \paren{1-\eta^{\overline{B}}}^{\frac{t-s}{\overline{B}}}.\]
\end{prop}

For simplicity, we shall denote 
$\Gamma = 2 \frac{1+\eta^{-\overline{B}}}{1-\eta^{\overline{B}}}$, 
$\gamma = \paren{1-\eta^{\overline{B}}}^{\frac{1}{\overline{B}}}$, 
and restate this theorem as 
\begin{align} \label{phi_m}
\left| [\Phi(t,s)]_{ij} - \oo{m} \right| \leq \Gamma \gamma^{t-s}.
\end{align}

This lemma will ensure that the distance between each agent's estimate and the average estimate
decreases geometrically with respect to the number of communication steps taken in the consensus stage. 
In particular, it gives the following bound on the distance between iterates $\qhik$, the outcome outcomes of the multi-step consensus stage \eqref{ch4_pg_2}, and their average, $\qbar{k} = \oo{m} \sumi \qhik$: 
\begin{align} \label{yeah2}
\norm{\qhik - \qbar{k}}
&= \norm{ \sumj \lijk \qjk - \oo{m} \qjk } \\
&\leq \sumj \left| \lijk - \oo{m} \right| \norm{\qjk} \nonumber \\
&\leq \Gamma \gamma^k \sumj \norm{\qjk}. \label{qbound}
\end{align}

\subsection{Formulation as an Inexact Method}

We now show that our method can be formulated as an inexact centralized proximal gradient method in the framework of \cite{SRB11}: 

\begin{prop} \label{l_inexactpg}
(Distributed proximal-gradient method as an inexact centralized proximal-gradient method)
Let $x_i^{( k )}$ and $y_i^{( k )}$ be iterates generated by Algorithm \eqref{ch4_pg}. Let $\xbar{k}=\oo{m} \sumi \xik$ and $\ybar{k}=\oo{m} \sumi \yik$ be the average iterates at iteration $k$. Then Algorithm \eqref{ch4_pg} can be written as 
\begin{align} \label{ch4_pg_inexact}
\begin{cases}
\xbar{k} &\in \prox_{h, \epsk}^\alpha \left\{ \ybar{k-1} - \alpha \brac{ \nabla g \paren{\ybar{k-1}} + \ek} \right\} \\
\ybar{k} &= \xbar{k} + \betak \paren{ \xbar{k} - \xbar{k-1} }
\end{cases}
\end{align}
where the error sequences $\{\ek\}_{k=1}^\infty$ and $\{\epsk\}_{k=1}^\infty$ satisfy
\begin{align}
\norm{\ek} \leq &\frac{L}{m} \sumi \norm{\yikm - \ybar{k-1}} \label{viga1}
\\
\epsk \leq &\frac{2G}{m} \sumi \norm{\qhik - \qbar{k}} \nonumber \\
&+ \oo{2 \alpha}\paren{\oo{m} \sumi \norm{\qhik - \qbar{k}}}^2. \label{viga2}
\end{align}

\begin{IEEEproof}
By taking the average of \eqref{ch4_pg_1}, we can see that 
\[\qbar{k} = \ybar{k-1} - \alpha (\nabla g(\ybar{k-1}) + \ek),\] 
where
\[\ek = \oo{m} \sumi \brac{\nabla g_i(\yikm) - \nabla g_i(\ybar{k-1})},\]
and therefore, due to the Lipschitz-continuity of the gradient of $g_i(x)$,
\[\norm{\ek} \leq \frac{L}{m} \sumi \norm{\yikm - \ybar{k-1}}.\]

Let
\begin{align*}
z^{( k )} &= \prox_h^\alpha \{ \qbar{k} \} = \argmin_x \left\{ h(x) + \oo{2 \alpha} \norm{x-\qbar{k}}^2 \right\}
\end{align*}
denote the result of the exact centralized proximal step. Then $\xbar{k} = \oo{m} \sumi \xik = \oo{m} \sumi \prox_h^\alpha \{ \qhik \}$, the result of the proximal step in the distributed method, can be seen as an approximation of $z^{( k )}$. 
We next relate $z^{( k )}$ and $\xbar{k}$ by formulating the latter as an inexact proximal step with error $\epsk$. 
A simple algebraic expansion gives
\begin{align*}
&h(\xbar{k}) + \oo{2 \alpha} \norm{\xbar{k}-\qbar{k}}^2 \\
\leq & h(z^{( k )}) + G_h\norm{\xbar{k}-z^{( k )}} + \oo{2 \alpha}\bigl\{\norm{z^{( k )}-\qbar{k}}^2 \\
&+ 2 \inn{z^{( k )}-\qbar{k}}{\xbar{k}-z^{( k )}} + \norm{\xbar{k}-z^{( k )}}^2 \bigr\} \\
=& \min_{z \in \ourdom}\left\{ h(z) + \oo{2 \alpha}\norm{z-\qbar{k}}^2 \right\} \\
&+ \norm{\xbar{k}-z^{( k )}} \paren{G_h + \oo{\alpha} \norm{z^{( k )}-\qbar{k}}} \\
&+ \oo{2 \alpha} \norm{\xbar{k}-z^{( k )}}^2,
\end{align*}
where in the inequality, we used the convexity of $h(x)$ and the bound on the subgradient $\partial h(\xbar{k})$ to obtain $h(\xbar{k}) \leq h(z^{( k )}) + G_h \norm{\xbar{k}-z^{( k )}} $; and in the equality, we used the fact that by definition, $z^{( k )}$ is the optimizer of $h(x) + \oo{2 \alpha} \norm{x-\qbar{k}}^2$. 

With this expression, we can write
\[\xbar{k} \in \prox_{h, \epsk}^\alpha \left\{ \qbar{k} \right\},\]
where
\begin{align*}
\epsk =& \norm{\xbar{k}-z^{( k )}} \paren{G_h + \oo{\alpha} \norm{z^{( k )}-\qbar{k}}} \\
&+ \oo{2 \alpha} \norm{\xbar{k}-z^{( k )}}^2.
\end{align*}

By definition, $z^{( k )} = \prox_h^\alpha \{ \qbar{k} \}$ also implies 
$\oo{\alpha} \paren{\qbar{k}-z^{( k )}} \in \partial h(z^{( k )})$, and therefore its norm is bounded by $G_h$. As a result,
\[\epsk \leq 2G_h \norm{\xbar{k}-z^{( k )}} + \oo{2 \alpha} \norm{\xbar{k}-z^{( k )}}^2.\]

Combined with the nonexpansiveness of the proximal operator (Proposition \ref{p_prox}(d)), 
\begin{align*}
\norm{\xbar{k} - z^{( k )}} 
&\leq \oo{m} \sumi \norm{\prox_h^\alpha\{\qhik\} - \prox_h^\alpha\{\qbar{k}\} } \\
&\leq \oo{m} \sumi \norm{\qhik - \qbar{k}},
\end{align*}
we arrive at the desired expression. 

\end{IEEEproof}
\end{prop}

This proposition shows that the two error sequences $\norm{\ek}$ and $\epsk$ have upper bounds in terms of $\oo{m} \sumi \norm{\yikm - \ybar{k-1}}$ and $\oo{m} \sumi \norm{\qhik - \qbar{k}}$, respectively, which are in turn controlled by the multi-step consensus stage. According to \cite[Proposition 2]{SRB11}, if $\{k\norm{\ek}\}$ and $\{k \sqrt{\epsk}\}$ are both summable, then the inexact proximal-gradient method exhibits the optimal exact convergence rate of $O(1/n^2)$. In the following sections, we shall see that this is indeed the case.

\subsection{Convergence Rate Analysis}

We next show that the sequences  $\{\norm{\ek}\}_{k=1}^\infty$ and $\{\epsk\}_{k=1}^\infty$  are bounded above by polynomial-geometric sequences. By Proposition \ref{l_polygeo}, this establishes that the sequences
$\{k\norm{\ek}\}$ and $\{k \sqrt{\epsk}\}$ are summable. We first present some useful recursive expressions of the iterates.

\begin{prop} (Recursive expressions of iterates) \label{p_recur_pg}

Let sequences $\{\xik\}_{k=1}^\infty, \{\yik\}_{k=1}^\infty, \{\qik\}_{k=1}^\infty$, $\{\qhik\}_{k=1}^\infty,$ $i=1,...,m$, be iterates generated by Algorithm \eqref{ch4_pg}. For every $k \geq 2$, we have
\begin{enumerate}[(a)]
\item $\sumi \norm{q_i^{k+1}}
\leq \sumi \norm{\qik} + \alpha m (G_g+G_h) + \sumi \norm{\xik - \xikm}$
\vspace*{0.1in}
\item $\sumi \norm{\xik - \xikm} 
\leq 2 m \Gamma \sum_{l=1}^{k-1} \gamma^l \sumi \norm{q_i^l} + (k-1) \alpha m (G_g+G_h)$
\vspace*{0.1in}
\item 
$\norm{\yik - \ybar{k}} \leq 4 \Gamma \gamma^k \sumi \norm{\qik} + 2 \Gamma \gamma^{k-1} \sumi \norm{\qikm}$
\end{enumerate} 
\end{prop}


The proof is given in the online appendix. These recursive expressions allows us to bound $\sumi \norm{\qik}$ with a second-order polynomial of $k$, as in the following lemma, whose proof is also given in the appendix.

\begin{lemma} \label{l_polybound_pg}
(Polynomial bound on $\sumj \norm{\qik}$) 

Let sequences $\{\xik\}_{k=1}^\infty,$ $\{\yik\}_{k=1}^\infty,$ $\{\qik\}_{k=1}^\infty,$ $\{\qhik\}_{k=1}^\infty,$ $i=1,...,m$, be generated by Algorithm \eqref{ch4_pg}. 
Then there exists scalars $C_q, C_q', C_q''$ such that for $k \geq 2$, 
\[\sumi \norm{\qik} \leq C_q + C_q' k + C_q'' k^2.\]
\end{lemma} 

We now apply Lemma \ref{l_polybound_pg} on the error sequences in \eqref{ch4_pg_inexact} to show that $\{k\norm{\ek}\}_{k=1}^\infty$ and $\{k \sqrt{\epsk}\}_{k=1}^\infty$ are polynomial-geometric sequences, thus summable:

\begin{lemma} 
(Summability of $\{k\norm{\ek}\}_{k=1}^\infty$ and $\{k \sqrt{\epsk}\}_{k=1}^\infty$)

In the formulation \eqref{ch4_pg_inexact}, where
\begin{align*}
\norm{\ek} \leq & \frac{L}{m} \sumi \norm{\yikm - \ybar{k-1}},
\\
\epsk \leq & \frac{2G_h}{m} \sumi \norm{\qhik - \qbar{k}} \\ 
&+ \oo{2 \alpha}\paren{\oo{m} \sumi \norm{\qhik - \qbar{k}}}^2,
\end{align*}
we have 
\begin{enumerate}[(a)]
\item $\sumk k\norm{\ek} < \infty$
\item $\sumk k \sqrt{\epsk} < \infty$
\end{enumerate}

\begin{IEEEproof}
In both cases, it suffices to show that the sequence is a polynomial-geometric sequence. The result then follows by Proposition \ref{l_polygeo}.

\begin{enumerate}[(a)]
\item 
By Proposition \ref{p_recur_pg}(c), 
\begin{align*}
&\oo{m} \sumi \norm{\yik - \ybar{k}} \\
\leq & 4 \Gamma \gamma^k \sumi \norm{\qik} + 2 \Gamma \gamma^{k-1} \sumi \norm{\qikm}
\end{align*}
and by Lemma \ref{l_polybound_pg}(a), 
$\sumi \norm{\qik} \leq C_q + C_q' k + C_q'' k^2$.
Therefore, 
\begin{align*}
k \norm{\ek} 
\leq &4 L \Gamma \gamma^k k \paren{C_q + C_q' k + C_q'' k^2} \\
&+ 2L \Gamma \gamma^{k-1} k \paren{C_q + C_q' (k-1) + C_q'' (k-1)^2},
\end{align*} 
which is a polynomial-geometric sequence.

\item
Recall \eqref{yeah2}, 
$\norm{\qhik - \qbar{k}} \leq \Gamma \gamma^k \sumj \norm{\qjk}$,
and Lemma \ref{l_polybound_pg}(a), 
$\sumj \norm{\qjk} \leq C_q + C_q' k + C_q'' k^2$.
Therefore,
\begin{align*}
\epsk \leq & 2 G_h \Gamma \gamma^k \paren{C_q + C_q' k + C_q'' k^2} \\
&+ \oo{2 \alpha} \brac{\Gamma \gamma^k \paren{C_q + C_q' k + C_q'' k^2}}^2.
\end{align*}
Using the fact that $\sqrt{a+b} \leq \sqrt{a} + \sqrt{b}$ for all nonnegative real numbers $a,b$, we have
\begin{align*}
\sqrt{\epsk} 
\leq & \sqrt{2 G_h \Gamma \gamma^k \paren{C_q + C_q' k + C_q'' k^2}} \\
& + \oo{\sqrt{2 \alpha}} \brac{\Gamma \gamma^k \paren{C_q + C_q' k + C_q'' k^2}} \\
\leq & \sqrt{2 G_h \Gamma} \sqrt{\gamma}^k \paren{\sqrt{C_q} + \sqrt{C_q'} k + \sqrt{C_q''} k} \\
& + \frac{\Gamma}{\sqrt{2 \alpha}} \gamma^k \paren{C_q + C_q' k + C_q'' k^2}
\end{align*}
where in the last line we used the fact that $\sqrt{k} \leq k$ for all $k \geq 1$. 
This is a polynomial-geometric sequence. 
Therefore, $k \sqrt{\epsk}$ is also a polynomial-geometric sequence.

\end{enumerate}
\end{IEEEproof}
\end{lemma}


Using the lemma above, we can establish the convergence rate of our distributed proximal-gradient method:

\begin{theorem} \label{t_pg}
(Convergence rate of the distributed proximal-gradient method with multi-step consensus)
Let $\{x_i^{( k )}\}_{k=1}^\infty$ be iterates generated by Algorithm \eqref{ch4_pg}, with a constant step size $\alpha\le 1/L$ where $L$ is the Lipschitz constant in Assumption \ref{a_func}. 
Let $\xbar{k}=\oo{m} \sumi \xik$ be the average iterate at iteration $k$. 
Then, for all $t \geq 1$, where $t$ is the total number of communication steps taken, we have 
\[f(\xbar{t}) - f(x^*) = O(1/t).\]

\begin{IEEEproof}
Since it takes $k$ communication steps to complete iteration $k$, the total number of communication steps required to execute iterations $1,...,n$ is
$\sumkn k = \frac{n(n+1)}{2}.$
In other words, after $t$ communication steps, the number of iterations completed is $n$, where $n$ is the greatest integer such that 
$\frac{n(n+1)}{2} = \frac{n^2 + n}{2} \leq t,$
or equivalently,
$n = \left \lceil \frac{-1+\sqrt{1+8t}}{2} \right \rceil.$
As a result, \[(n+1)^2 \geq \paren{\frac{-1+\sqrt{1+8t}}{2}}^2 = \frac{2+8t-2\sqrt{1+8t}}{4},\]
and thus, \[f(\xbar{t}) - f(x^*) \leq \frac{D}{(n+1)^2} \leq \frac{2D}{4t+1-\sqrt{1+8t}} = O(1/t).\]
\end{IEEEproof}

\end{theorem}

Although this theorem is stated in terms of $\xbar{t}$, it could also lead to a bound on $f(x_i^{(t)}) - f(x^*)$, using the gradient bound, nonexpansiveness of the proximal operator, and \eqref{qbound}. Also, note that the results above hold for the fast distributed gradient method (where the objective functions are differentiable), which is clear by simply setting $h(x)=0, G_h = 0$ and $\epsk = 0$.


\subsection{Beyond O($1/t$)}

We have thus shown that taking $k$ communication steps in the $k$-th iteration of Algorithm \eqref{ch4_pg} results in the summability of error sequences $\{k\norm{\ek}\}_{k=1}^\infty$ and $\{k \sqrt{\epsk}\}_{k=1}^\infty$. A natural question arises: can we do better? In particular, will the error sequences still converge if we took less than $k$ communication steps in the $k$-th iteration? We address this question in this section.

Let $s_k$ be the number of communication steps taken in the multi-step consensus stage at iteration $k$. In our method presented earlier, $s_k = k$. We wish to explore smaller choices of $s_k$ that still preserves the guarantee for exact convergence. 

With $s_k$, we see that \eqref{yeah2} can be written as
\[\norm{\qhik - \qbar{k}} \leq \Gamma \gamma^{s_k} \sumj \norm{\qjk}.\]
Therefore, Proposition \ref{p_recur_pg}(b) becomes
\begin{align*}
&\sumi \norm{\xik - \xikm} \\
\leq &\sum_{l=1}^{k-1} 2 m \Gamma \gamma^{s_l} \sumj \norm{q_j^l} + ( k-1 ) \alpha m (G_g+G_h)
\end{align*}

As a result, if we have the equivalent of Proposition \ref{l_polygeo} for $s_k$, i.e., if $\sumkz k^N \gamma^{s_k} < \infty$ for any given $\gamma \in (0,1)$ and nonnegative integer $N$, then Lemma \ref{l_polybound_pg} would hold, and so would Theorem \ref{t_pg}. 

Since $\sumkz k^a < \infty$ for $a<-1$, a sufficient condition for the above is $\gamma^{s_k} < k^{-N-1},$ or equivalently, $s_k > \frac{-N-1}{\log \gamma}\log k.$ 
This is at the order of $O(\log k)$, which is smaller than our previous choice of $s_k = k = O^{( k )}$. The hidden constant, $\frac{-N-1}{\log \gamma}$, depends on $N$ and $\gamma$. In our case, we only require this condition to hold up to $N = 3$. Therefore, if $\gamma$ is known, by choosing 
$ s_k = \left\lceil \frac{4}{- \log \gamma} \log (k+1) \right\rceil$, 
the distributed proximal-gradient method is guaranteed to converge with rate $O(1/n^2)$, where $n$ is the iteration number. 

The time it takes to complete iterations $1,...,n$, which we denote by $T(n)$, is then \[T(n) = \sumkn s_k = O(n \log n - n)\]
since $\int \log x dx = x (\log x - 1)$. Unfortunately, $n \log n = \log n \cdot e^{\log n}$ has no explicit inverse expression \cite{Cor96}. Therefore, we can only express the convergence rate as 
\[f(\xbar{t}) - f(x^*) = O(1/(T^{-1}(t))^2)\]
which we know to be better than $O(1/t)$, since $T(n)$ is bounded above by $O(n^2)$.

In closing, we remark that the improved choice of $s_k$ above requires the knowledge of $\gamma$, which may not be readily available if detailed information or performance guarantees of the communication network is unknown. In such cases, the method could still be implemented with $s_k = k$.

\section{Numerical Experiments}

\begin{figure*}[!ht]
  \centering
  \includegraphics[width=0.65\textwidth]{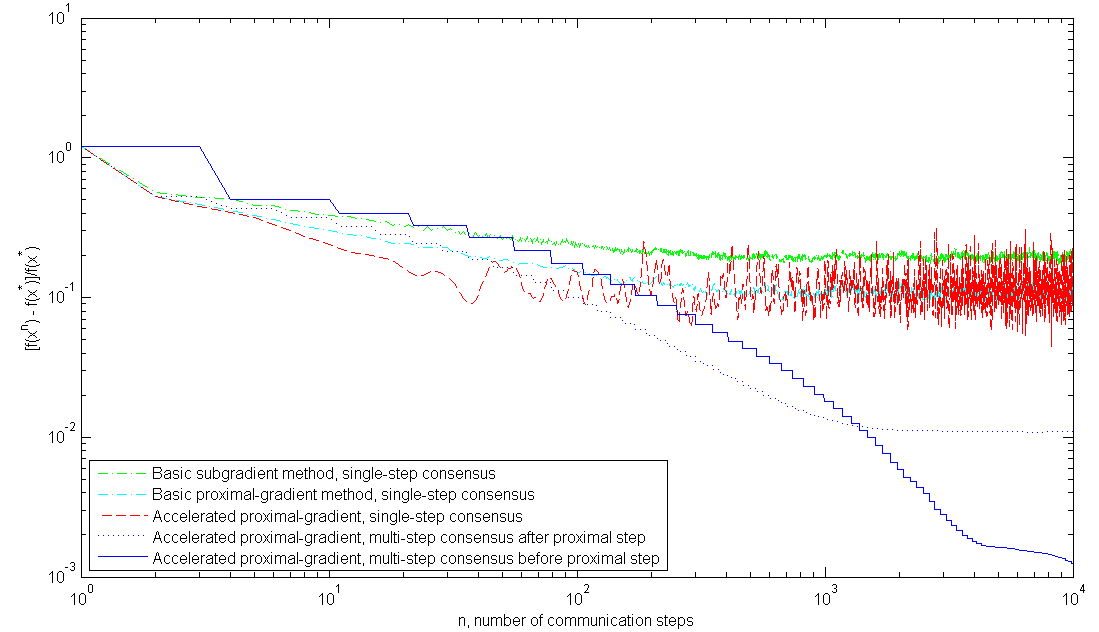}
  \caption{Performance comparison for various distributed methods} \label{fig:exp1}
\end{figure*}

Our theoretical findings are verified with numerical experiments on a machine learning task using 20 Newsgroups \cite{NG1,NG2}, a benchmark dataset for text categorization. It consists of about 20,000 news articles, evenly chosen among 20 topics. The task is to perform $L_1$-regularized logistic regression on the training data, so as to learn the classification model for a chosen topic label. Specifically, we wish to minimize
\[f(x) = \oo{N} \sum_{j=1}^N \log \paren{1+\exp (-b_j \inn{a_j}{x})} + \lambda \norm{x}_1, \]
where $N$ is the total number of news articles, $a_j$ is the 8615-dimensional feature vector of article $j$, and $b_j$ is its the label for the chosen topic, which is equal to $1$ if this article belongs to the topic, and $-1$ otherwise. $x$ contains parameters of the classification model that we wish to learn, and $f(x)$ is its corresponding regularized loss function. 

We distribute the training data across a network of $m=10$ data centers, each with 1129 samples. Thus, each data center has the following private objective function:
\[f_i(x_i) = \oo{|N_i|} \sum_{j \in N_i} \log \paren{1+\exp (-b_j \inn{a_j}{x_i})} + \lambda \norm{x_i}_1 \]
where $N_i$ is the subset of data at center $i$, and $x_i, i=1,...,m$, is its local estimate of the global classification model. 
In each communication step, a weight matrix is randomly chosen from a pool of 10 weight matrices generated from connected random graphs. All weight matrices satisfy Assumption \ref{a_network}. 

To demonstrate the effect of using multiple communication steps after the gradient step in our method, we compare it with the following methods:
\begin{itemize}
\item The \emph{basic} subgradient method with \emph{single-step} consensus in \cite{NeO09}:
\begin{small}

\begin{align} \label{ch3_basic_subgrad}
	\begin{cases}
	\xik &= \wikm - \ak \paren{\nabla g_i(\wikm) + z_h(\wikm)} \\
	\wik &= \sumj a_{ij}^{( k )} \xjk
	\end{cases}
\end{align}
\end{small}
where $z_h(\wikm) \in \partial h(\wikm)$ and $\brac{a_{ij}^{( k )}}$ is the randomly-chosen weight matrix.

\item The \emph{basic} proximal-gradient method with \emph{single-step} consensus, similar to that of \cite{NeO09}:
\begin{small}
\begin{align} \label{ch3_basic}
	\begin{cases}
	\xik &= \prox_h^\alpha\{ \wikm - \ak \nabla g_i(\wikm) \} \\
	\wik &= \sumj a_{ij}^{( k )} \xjk
	\end{cases}
\end{align}
\end{small}
where $\brac{a_{ij}^{( k )}}$ is the randomly-chosen weight matrix.

\item The \emph{accelerated} proximal-gradient method with \emph{single-step} consensus:
\begin{small}
\begin{align} \label{ch3_acce}
	\begin{cases}
	\xik &= \prox_h^\alpha\{ \wikm - \ak \nabla g_i(\wikm) \} \\
	\yik &= \xik + \betak(\xik - \xikm) \\
	\wik &= \sumj a_{ij}^{( k )} \yjk
	\end{cases}
\end{align}
\end{small}

\item The \emph{accelerated} proximal-gradient method with \emph{multi-step} consensus which is not inserted between the gradient and proximal steps, but instead performed only after the proximal step:
\begin{small}
\begin{align} \label{ch4_pg_bad}
\begin{cases}
	\xik &= \prox_h^\alpha\{ \wikm - \ak \nabla g_i(\wikm) \} \\
	\yik &= \xik + \betak \paren{\xik - \xikm} \\
	\wik &= \sumj \lijk \yjk \\
\end{cases}
\end{align}
\end{small}
where $\brac{\lijk}$ is the product of $k$ weight matrices randomly drawn from the pool of 10 weight matrices.

\end{itemize} 

Figure \ref{fig:exp1} shows convergence rate results for each method. 
It is clear from the figure that Algorithm \eqref{ch3_basic_subgrad} converges to an error neighborhood at rate $O(1/t)$, as shown in \cite{NeO09}. 
Algorithms \eqref{ch3_basic} and \eqref{ch3_acce} also converge to an error neighborhood, but the latter exhibits more oscillation than the basic methods. 
Algorithm \eqref{ch4_pg_bad} converges with rate $O(1/t)$, but only to an error neighborhood instead of achieving exact convergence. This highlights the importance of having the consensus step before the proximal step instead of after it. 
Finally, our accelerated multi-step method attains exact convergence with rate $O(1/t)$, outperforming all others.

\section{Conclusion and Future Work}

We presented a distributed proximal-gradient method that solves for the optimum of the average of convex functions, each having a distinct differentiable component and a common nondifferentiable component. The method uses multiple communication steps and Nesterov's acceleration technique. We established the convergence rate of this method as $O(1/t)$ (where $t$ is the total number of communication steps), superior to most existing distributed methods.

Several questions remain open for future work. First, it would be useful to generalize the result for the case where the nondifferentiable functions $h_i(x)$ are distinct. Secondly, it is of interest to determine the condition under which the accelerated single-step proximal-gradient method \eqref{ch3_acce} converges, and compare its performance with our multi-step consensus method. Last but not least, it would be useful to obtain a lower bound on the convergence rate of distributed first-order methods under our current framework.

\bibliographystyle{IEEEtran}
\bibliography{distributed_proxgrad}

\begin{thebibliography}{10}
\providecommand{\url}[1]{#1}
\csname url@samestyle\endcsname
\providecommand{\newblock}{\relax}
\providecommand{\bibinfo}[2]{#2}
\providecommand{\BIBentrySTDinterwordspacing}{\spaceskip=0pt\relax}
\providecommand{\BIBentryALTinterwordstretchfactor}{4}
\providecommand{\BIBentryALTinterwordspacing}{\spaceskip=\fontdimen2\font plus
\BIBentryALTinterwordstretchfactor\fontdimen3\font minus
  \fontdimen4\font\relax}
\providecommand{\BIBforeignlanguage}[2]{{%
\expandafter\ifx\csname l@#1\endcsname\relax
\typeout{** WARNING: IEEEtran.bst: No hyphenation pattern has been}%
\typeout{** loaded for the language `#1'. Using the pattern for}%
\typeout{** the default language instead.}%
\else
\language=\csname l@#1\endcsname
\fi
#2}}
\providecommand{\BIBdecl}{\relax}
\BIBdecl

\bibitem{NeO09}
A.~Nedic and A.~Ozdaglar, ``Distributed subgradient methods for multi-agent
  optimization,'' in \emph{LIDS report 2755, IEEE Transactions on Automatic
  Control}, vol.~54, no.~1, 2009, pp. 48--61.

\bibitem{DAW11}
J.~Duchi, A.~Agarwal, and M.~Wainwright, ``Dual averaging for distributed
  optimization: Convergence and network scaling,'' \emph{IEEE Transactions on
  Automatic Control}, 2012.

\bibitem{JXM11}
D.~Jakovetic, J.~Xavier, and J.~M.~F. Moura, ``Fast distributed gradient
  methods,'' \emph{\url{arXiv:1112.2972v1}}, 2011.

\bibitem{SRB11}
M.~Schmidt, N.~L. Roux, and F.~Bach, ``Convergence rates of inexact
  proximal-gradient methods for convex optimization,'' \emph{CoRR}, vol.
  abs/1109.2415, 2011.

\bibitem{Tsi84}
J.~N. Tsitsiklis, ``Problems in decentralized decision making and
  computation,'' Ph.D. dissertation, Department of EECS, MIT, 1984.

\bibitem{TBA86}
J.~N. Tsitsiklis, D.~P. Bertsekas, and M.~Athans, ``Distributed asynchronous
  deterministic and stochastic gradient optimization algorithms,'' \emph{IEEE
  Transactions on Automatic Control}, vol.~31, no.~9, pp. 803--812, 1986.

\bibitem{Jad03}
A.~Jadbabaie, J.~Lin, and S.~Morse, ``Coordination of groups of mobile
  autonomous agents using nearest neighbor rules,'' \emph{IEEE Transactions on
  Automatic Control}, vol.~48, no.~6, pp. 988--1001, 2003.

\bibitem{Olf04}
R.~Olfati-Saber and R.~Murray, ``Consensus problems in networks of agents with
  switching topology and time-delays,'' \emph{IEEE Transactions on Automatic
  Control}, vol.~49, no.~9, pp. 1520--1533, 2004.

\bibitem{Boy05}
S.~Boyd, A.~Ghosh, B.~Prabhakar, and D.~Shah, ``Gossip algorithms: Design,
  analysis, and applications,'' \emph{Proceedings of IEEE INFOCOM}, 2005.

\bibitem{Ols06}
A.~Olshevsky and J.~Tsitsiklis, ``Convergence rates in distributed consensus
  and averaging,'' \emph{Proceedings of the 45th IEEE Conference on Decision
  and Control}, 2006.

\bibitem{Joh08}
B.~Johansson, T.~Keviczky, M.~Johansson, and K.~Johansson, ``Subgradient
  methods and consensus algorithms for solving convex optimization problems,''
  \emph{Proceedings of the 47th IEEE Conference on Decision and Control}, p.
  4185–4190, 2008.

\bibitem{Ram09}
S.~S. Ram, A.~Nedic, and V.~V. Veeravalli, ``Asynchronous gossip algorithms for
  stochastic optimization,'' \emph{Proceedings of the 48th IEEE Conference on
  Decision and Control}, pp. 3581--3586, 2009.

\bibitem{NOP10}
A.~Nedic, A.~Ozdaglar, and P.~A. Parrilo, ``Constrained consensus and
  optimization in multi-agent networks,'' \emph{IEEE Transactions on Automatic
  Control}, vol.~55, no.~4, 2010.

\bibitem{LoO11}
I.~Lobel and A.~Ozdaglar, ``Distributed subgradient methods for convex
  optimization over random networks,'' \emph{IEEE Transactions on Automatic
  Control}, vol.~56, no.~6, pp. 1291--1306, 2011.

\bibitem{Zhu12}
M.~Zhu and S.~Mart{\'\i}nez, ``On distributed constrained formation control in
  operator-vehicle adversarial networks,'' \emph{Automatica}, submitted, 2012.

\bibitem{Boy10}
S.~Boyd, N.~Parikh, E.~Chu, B.~Peleato, and J.~Eckstein, ``Distributed
  optimization and statistical learning via the alternating direction method of
  multipliers,'' \emph{Foundations and Trends in Machine Learning}, vol.~3,
  no.~1, 2011.

\bibitem{Wei12}
E.~Wei and A.~Ozdaglar, ``Distributed alternating direction method of
  multipliers.''

\bibitem{Roc76}
R.~T. Rockafellar, ``Monotone operators and the proximal point algorithm,''
  \emph{SIAM Journal on Control and Optimization}, vol.~14, no.~5, pp.
  877--898, 1976.

\bibitem{BeT09}
A.~Beck and M.~Teboulle, ``A fast iterative shrinkage-thresholding algorithm
  for linear inverse problems,'' \emph{SIAM Journal on Imaging Sciences},
  vol.~2, pp. 183--202, March 2009.

\bibitem{BeT10}
------, ``Gradient-based algorithms with applications in signal recovery
  problems,'' in \emph{Convex Optimization in Signal Processing and
  Communications}, D.~P. Palomar and Y.~C. Eldar, Eds.\hskip 1em plus 0.5em
  minus 0.4em\relax Cambridge University Press, 2010.

\bibitem{BHO05}
V.~D. Blondel, J.~M. Hendrickx, A.~Olshevsky, and J.~N. Tsitsiklis,
  ``Convergence in multiagent coordination, consensus, and flocking,''
  \emph{Proceedings of the Joint 44th IEEE Conference on Decision and Control
  and European Control Conference (CDC-ECC'05)}, 2005.

\bibitem{Cor96}
R.~M. Corless, G.~H. Gonnet, D.~E.~G. Hare, D.~J. Jeffrey, and D.~E. Knuth,
  ``On the lambert w function,'' in \emph{Advances in Computational
  Mathematics}, 1996, pp. 329--359.

\bibitem{NG1}
K.~Lang, ``Newsweeder: Learning to filter netnews,'' in \emph{Proceedings of
  the Twelfth International Conference on Machine Learning}, 1995, pp.
  331--339.

\bibitem{NG2}
J.~Rennie, ``20 newsgroups,''
  \url{http://people.csail.mit.edu/jrennie/20Newsgroups/}.

\end{thebibliography}

\newpage
\section*{Appendix}

\subsection*{Proof of Proposition \ref{p_recur_pg}}

Throughout this proof, let $\beta_k = \betak$ for simplicity. Moreover, it is useful to note that, by Proposition \ref{p_prox}, \eqref{ch4_pg_3} could be written as  
\begin{align} \label{ch4_pg_3a}
\xik = \qhik - \alpha \zik, \mbox{ where } \zik \in \partial h\paren{\xik}.
\end{align} 
Since $h$ has bounded subgradients, this also implies
\begin{align} \label{ch4_pg_3b}
\norm{ \xik - \qhik } \leq \alpha G_h.
\end{align}

\begin{enumerate}[(a)]
\item 
Taking norm of \eqref{ch4_pg_1} and summing over $i$, we have
\begin{align} \label{zena0}
\sumi \norm{\qik} &=  \sumi \norm{\yikm - \alpha \nabla g_i(\yikm)} \nonumber \\
&\leq \sumi \norm{\yikm} + \alpha m G_g,
\end{align}
where we used the gradient bound in Assumption \ref{a_func}.

According to \eqref{ch4_pg_4}, we have \[\yikm = \xikm + \betakm (\xikm - x_i^{( k-2 )}),\] and by \eqref{ch4_pg_3b}, we have \[\norm{\xikm} - \norm{\qhikm} \leq \norm{\xikm-\qhikm} \leq \alpha G_h.\] Therefore,
\begin{align} \label{zena2}
&\norm{\yikm} 
\leq \norm{\qhikm} + \alpha G_h + \betakm \norm{\xikm - x_i^{( k-2 )}}.
\end{align}

Next, we use \eqref{ch4_pg_2}, which states that $\qhikm = \sumj \lijkhatm \qjkm$ is a convex combination of $\{\qjkm\}_{j=1}^m$, so
\begin{align} \label{zena3}
\sumi \norm{\qhikm} \leq \sumi \norm{\qikm}.
\end{align}

Substituting \eqref{zena2}-\eqref{zena3} back in \eqref{zena0}, we have 
\begin{align*}
\sumi \norm{\qik}
\leq & \sumi \norm{\qikm} + \alpha m (G_g + G_h) \\
&+ \betakm \sumi \norm{\xikm - x_i^{( k-2 )}}.
\end{align*}

Finally, we omit $\betakm \leq 1$, and increment the indices by $1$ so that the expression is applicable to $k \geq 2$.


\item Starting with \eqref{ch4_pg_3a} and applying \eqref{ch4_pg_2}, \eqref{ch4_pg_1}, \eqref{ch4_pg_4} in order, we have
\begin{align*}
\xik =& \qhik - \alpha \zik 
= \sumj \lijkhatm \qjk - \alpha \zik \\
=& \sumj \lijkhatm \brac{\yjkm - \alpha \nabla g_j\paren{\yjkm}} - \alpha \zik \\
=& \sumj \lijkhatm \brac{\xjkm + \betakm\paren{\xjkm-x_j^{( k-2 )}} }\\ 
&- \sumj \lijkhatm \alpha \nabla g_j\paren{\yjkm} - \alpha \zik.
\end{align*}

Subtracting $\xikm$ from the previous expression and taking the sum of the norm, we have
\begin{align} \label{xero}
&\sumi \norm{\xik - \xikm}  \nonumber \\
\leq & \sumi \sumj \lijkhatm \norm{\xjkm - \xikm} \nonumber \\
&+ \betakm \sum_{j=1}^m \norm{x_j^{( k-1 )}-x_j^{( k-2 )}} + \alpha m (G_g+G_h),
\end{align}
where we used the convexity of the norm operator along with the fact that $\sumi \lijkhatm = 1$.

Now consider $\sumi \sum_{j=1}^m \lijkhatm \norm{x_j^{k-1} - \xikm}$ in the expression above. By the nonexpansiveness of the proximal operator, we have 
$\norm{\xjkm - \xikm} \leq \norm{\qhjkm - \qhikm}. $
Using the fact that $\brac{\sumi \lijkhatm}$ is doubly stochastic, we have
\begin{align*}
&\sumi \sumj \lijkhatm \norm{\qhjkm - \qhikm} \\
\leq & \sumi \sumj \lijkhatm \paren{\norm{\qhikm-\qbar{k-1}} +\norm{\qhjkm-\qbar{k-1}} }\\
= & 2 \sumi \norm{\qhikm - \qbar{k-1}}.
\end{align*}

The right-hand side can in turn be bounded with \eqref{yeah2}. 
As a result,
\begin{align} \label{yeah3}
&\sumi \sumj \lijkhatm \norm{\xjkm - \xikm} \nonumber\\
\leq & 2 m \Gamma \gamma^{k-1} \sumj \norm{\qjkm}.
\end{align}

Substituting this back to \eqref{xero},
\begin{align*}
&\sumi \norm{\xik - \xikm} \\
\leq & 2 m \Gamma \gamma^{k-1} \sumj \norm{\qjkm} \\
&+ \betakm \sumi \norm{x_i^{( k-1 )}-x_i^{( k-2 )}} + \alpha m (G_g+G_h) \\
\leq & \sum_{l=1}^{k-1} \paren{2 m \Gamma \gamma^l \sumj \norm{q_j^l} + \alpha m (G_g+G_h) }
\end{align*}
where the final line is due to recursion, and omitting $\beta_l \leq 1$ for $l > 1$ while using $\beta_1 = 0$ to eliminate the tailing term $\sumi \norm{x_i^1-x_i^0}$. This is the desired expression.

\item
By \eqref{ch4_pg_4},
\[
\yik - \ybar{k} = (1+\beta_k) (\xik-\xbar{k}) - \beta_k (\xikm-\xbar{k-1}).
\]
Note also that 
$\sumi \norm{\xik-\xbar{k}}
\leq \oo{m} \sumi \sumj \norm{\xik - \xjk}
\leq  2 \Gamma \gamma^k \sumj \norm{\qjk}$
by a similar reasoning as that of \eqref{yeah3}. Therefore, 
\begin{align*}
&\norm{\yik - \ybar{k}} \\
\leq & (1+\beta_k) \norm{\xik-\xbar{k}} + \beta_k \norm{\xikm-\xbar{k-1}} \\
\leq & (1+\beta_k) 2 \Gamma \gamma^k \sumj \norm{\qjk} + \beta_k 2 \Gamma \gamma^{k-1} \sumj \norm{\qjkm}.
\end{align*}
Omitting $\beta_k < 1$ gives statement (c). 

\end{enumerate}

\subsection*{Proof of Lemma \ref{l_polybound_pg}}

We proceed by induction on $k$. First, we show that the result holds for $k=2$ by choosing $C_q = \sumi \norm{q_i^{(2)}}$. It suffices to show that, given the initial points $y_j^0$, $\sumj \norm{q_j^{(2)}}$ is bounded. 

Indeed, by \eqref{zena0},
\[\sumi \norm{q_i^{(1)}} \leq \sumi \norm{y_i^{(0)}} + \alpha m G_g < \infty\]
and 
\begin{align*}
\sumi \norm{q_i^{(2)}} 
&\leq \sumj \norm{y_i^{(1)}} + \alpha m G_g 
= \sumi \norm{x_i^{(1)}} + \alpha m G_g \\
&\leq \sumi \norm{q_i^{(1)}} + \alpha m (G_g+G_h) 
< \infty
\end{align*}
where the first line is due to the fact that $\beta_1 = 0$ so $y_i^{(1)} = x_i^{(1)}$, and the second line is because of \eqref{ch4_pg_3a} and \eqref{zena3}. Therefore, $C_q = \sumi \norm{q_i^{(2)}} < \infty$ is a valid choice. 

Now suppose the result holds for some positive integer $k \geq 2$. We show that it also holds for $k+1$.

Substituting the induction hypothesis for $k$ into Proposition \ref{p_recur_pg}(b), we have
\begin{align*}
&\sumi \norm{\xik - \xikm} \\
\leq & 2 m \Gamma \sum_{l=1}^{k-1} \gamma^l \paren{C_q + C_q' l + C_q'' l^2} + (k-1) \alpha m (G_g+G_h)
\end{align*}

By Proposition \ref{l_polygeo} and expression \eqref{sngamma}, there exists constants $S_0^\gamma, S_1^\gamma, S_2^\gamma$ such that
\[
\sum_{l=0}^{\infty} \gamma^l \paren{C_q + C_q' l + C_q'' l^2} 
\leq C_q S_0^\gamma + C_q' S_1^\gamma + C_q'' S_2^\gamma.
\]

Proposition \ref{p_recur_pg}(a) and the induction hypothesis then gives us
\begin{align*}
\sumi \norm{q_i^{k+1}}
\leq &
C_q + C_q' k + C_q'' k^2 + \alpha m (G_g+G_h) \\
&+ 2 m \Gamma \paren{C_q S_0^\gamma + C_q' S_1^\gamma + C_q'' S_2^\gamma} \\
&+ (k-1) \alpha m (G_g+G_h)
\end{align*}

Comparing coefficients, we see that the right-hand side can be bounded by $C_q + C_q' (k+1) + C_q''(k+1)^2$  
if 
$\alpha m (G_g+G_h) \leq 2 C_q''$
for the coefficient of $k$, and
$2 m \Gamma \paren{C_q S_0^\gamma + C_q' S_1^\gamma + C_q'' S_2^\gamma} \leq C_q' + C_q''$
for the constant coefficient. Therefore, the induction hypothesis holds for $k+1$ if we take 
\begin{align*}
C_q &= \sumi \norm{q_i^2}, \\
C_q' &= \frac{2 m \Gamma C_q S_0^\gamma + (2 m \Gamma S_2^\gamma - 1) C_q''}{2 m \Gamma S_1^\gamma - 1}, \\
C_q'' &= \oo{2} \alpha m (G_g+G_h).
\end{align*}

\end{document}